\documentclass[preprint,showkeys,aps,prb]{revtex4}
\usepackage[dvips]{graphicx}
\begin{document}

\title{
Computational Implementation of Maxwell's Knudsen-Gas Demon and 
Its Extension to a Two-Dimensional Soft-Disk Fluid
}

\author{
William Graham Hoover and Carol Griswold Hoover                           \\
Ruby Valley Research Institute                  \\
601 Highway Contract 60                         \\
Ruby Valley, Nevada 89833                       \\
}

\date{\today}


\vspace{0.1cm}

\begin{abstract}
An interesting preprint by Puru Gujrati, ``Maxwell's Demon Must Remain Subservient
to Clausius' Statement'' [ of the Second Law of Thermodynamics ], traces the
development and application of Maxwell's Demon. He argues against the Demon on
thermodynamic grounds. Gujrati introduces and uses his own version of a generalized
thermodynamics in his criticism of the Demon. The complexity of his paper and the
lack of any accompanying numerical work piqued our curiousity.  The internet provides
well over two million ``hits'' on the subject of ``Maxwell's Demon''.  There are also
hundreds of images of the Demon, superimposed upon a container of gas or liquid.
However, there is not so much along the lines of simulations of the Demonic process.
Accordingly, we thought it useful to write and execute relatively simple FORTRAN
programs designed to implement Maxwell's low-density model and to develop its
replacement with global Nos\'e-Hoover or local purely-Newtonian thermal controls. These
simulations illustrate the entropy decreases associated with all three types of Demons. 
\end{abstract}

\maketitle

\section{Introduction and Goal}

Maxwell's Demon is a familiar example of a dynamical system designed to violate the
Second Law of Thermodynamics.  By deciding whether or not to reflect particles nearing
a movable barrier, open or shut based on kinetic energy or species, Maxwell's Demon is
able to generate a temperature or concentration difference without violating any mechanical
laws. Gujrati\cite{b1} provides extensive references to the Demon problem in his arXiv
preprint.  Wikipedia is another useful source reference.

We have two goals here: first, we implement Maxwell's original idea, quite suitable for
informal laptop simulations.  This implementation pointed out a need for extending
Maxwell's idea to dense fluids. That work, our second goal, is a surprisingly tall
order. We achieve it here, first by introducing two Nos\'e-Hoover Demons, and next, a
Soft-Disk Dense-Fluid Demon. We believe that such numerical
demonstrations are superior to theoretical armchair discussions and speculations. Our
hope is that the reader will find our implementations interesting, perhaps motivating
further improvements extending the reach of Maxwell's Demon idea.  We begin by modelling
entropy reduction with a two-dimensional (collisionless) Knudsen gas. That approach fails
for dense matter, as Gujarti stresses. We extend the scope of Demons by formulating new
ones based on contemporary Nos\'e-Hoover dynamics as well as a Newtonian soft-disk
dense-fluid Demon with a bit of memory.

\begin{figure}
\includegraphics[width=2.5 in,angle=-0.]{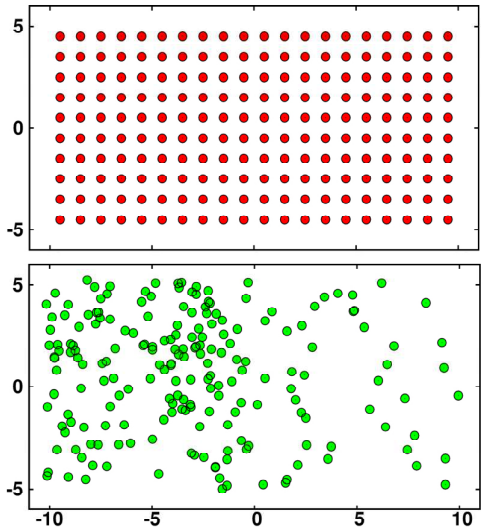}
\caption{
Configuration of 200 particles at time = 0 (above) and at time = 10,000
(below). Here the parameter of borderline energy between reflection and
penetration is 1/10 where $E/N$ is unity. Energetic particles go ``right'' while the
less-energetic particles are reflected to the ``left''.
}
\end{figure}

\section{Entropy decrease model: a two-dimensional Knudsen gas}

For convenience we choose 200 particles in the square-lattice configuration shown in
{\bf Figure 1}. The particles inhabit a 20x10 container centered at the origin. The
boundary conditions at the four walls are enforced, using cubic repulsive forces, by
four lines of code:
\begin{verbatim}
if (x(i).gt.+10) fx(i) = -800*(x(i)-10)^3 
if (x(i).lt.-10) fx(i) = -800*(10+x(i))^3
if (y(i).gt.+ 5) fy(i) = -800*(y(i)- 5)^3
if (y(i).lt.- 5) fy(i) = -800*( 5+y(i))^3
\end{verbatim}
The quartic boundary potential generating these forces was selected to resemble the
short-ranged purely-repulsive soft-disk potential $\phi(r<1) = 100(1 - r^2)^4$.
That pair potential has been used to determine the number-dependence of the shear
viscosity in relatively large two-dimensional simulations, with as many as
$N = 514\times 514$ particles\cite{b2}. To begin we consider a confined dynamics
without any particle-particle interaction, a ``Knudsen gas''. Later we will take
up the extension of Maxwell's goal to the case where particles interact with the
simple repulsive soft-disk pair potential. For an introduction to the Maxwell
problem we consider dynamics solely subject to the four reflecting boundaries at
$x = \pm 10$ and $y = \pm 5$.

\begin{figure}
``\includegraphics[width=2.1 in,angle=-90.]{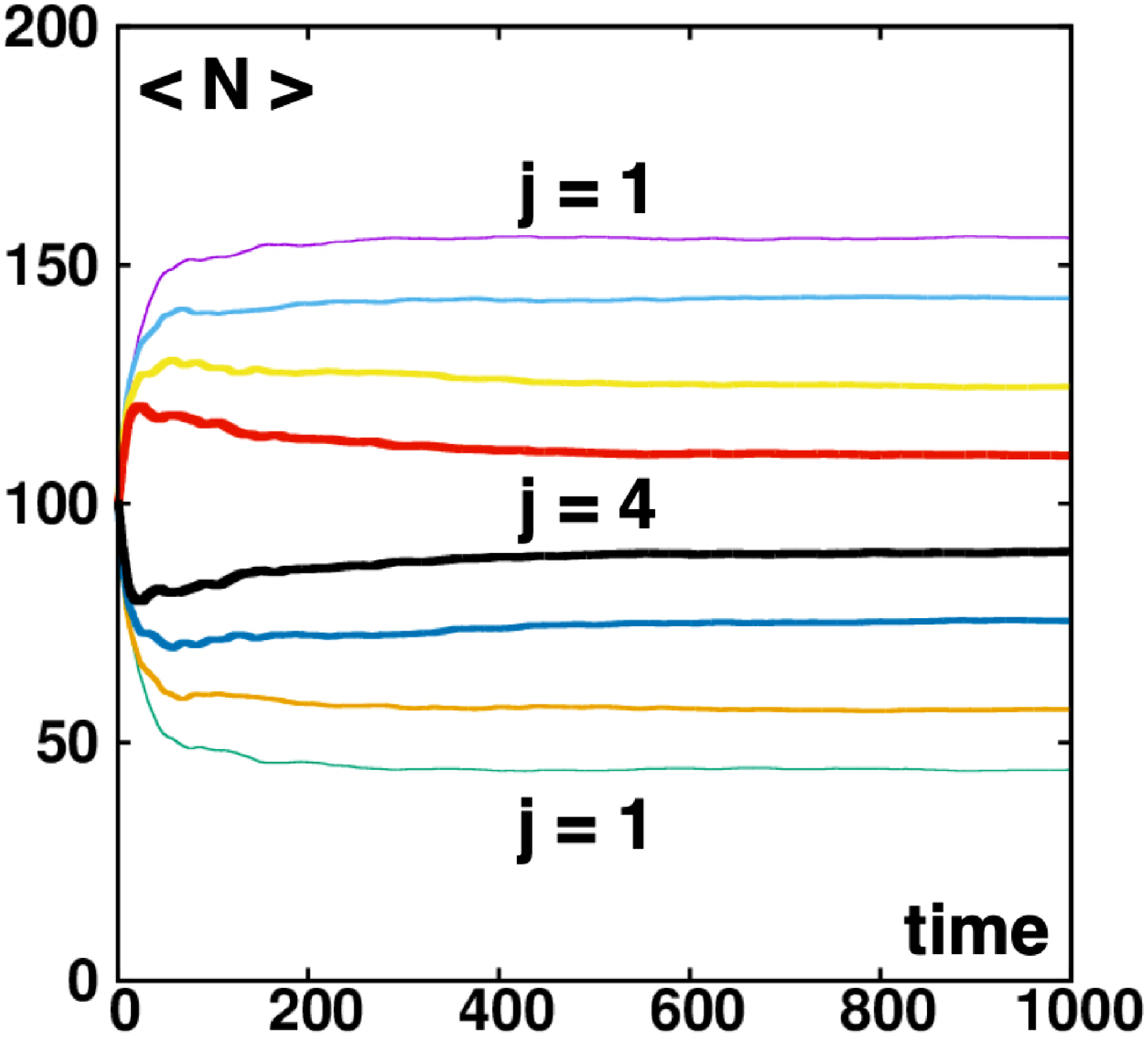}
\caption{
Time averaged hot (below) and cold (above) populations $\langle \ N \ \rangle$
with four mirror settings, $(j/8)$ for $j$ = 1, 2, 3, and 4. The mean particle
energy is unity. These four Knudsen gas simulations exclude interactions between
pairs of particles. Instead all particles are confined by repulsive quartic
potentials at $x=\pm 10$ and $y=\pm 5$. Any particle crossing the center line
$x=0$ is transmitted or reflected according to its total energy (same as its
kinetic temperature within the $20\times 10$ rectangle) as explained in the text.
}
\end{figure}

We choose our initial velocities $\{ \ p_x,p_y \ \}$ randomly in the interval
$\pm 0.5$: $\{ \ r - 0.5\  \}$. The FORTRAN routine ${\tt random}\_{\tt number(r)}$
is a convenient choice. Next, the particle velocities are scaled so as to give an
initial kinetic energy per particle of unity. For a well-posed problem all we
need do subsequently is to specify the workings of the Demon. He does not
disturb mass or energy conservation. His only job is to break the thermal
symmetry between the two halves of the system, creating a temperature difference
across the midline $x=0$. He can do this by implementing four lines of FORTRAN
applied whenever the product of a current $x_i(t)$ coordinate at time $t$ and
its previous value $x_i(t-dt)$ at time $t-dt$ is negative, indicating a crossing
of the midline at $x=0$:
$$
(p_{x_i}^2 + p_{y_i}^2)/2<(1/10)(E/N)\longrightarrow
 x_i = -\sqrt{x_i^2} \ ; \ p_{x_i} = -\sqrt{p_{x_i}^2} \ ;
$$
$$
(p_{x_i}^2 + p_{y_i}^2)/2>(1/10)(E/N)\longrightarrow
 x_i = +\sqrt{x_i^2} \ ; \ p_{x_i} = +\sqrt{p_{x_i}^2} \ .
$$

{\bf Figure 1} displays both the initial configuration and another much later, at a time
of $10^4$, for the particular reflecting borderline choice $E/10N$. Notice that
the reflection of left-to-right travel results in a higher density and lower
temperature in the left half of the system. {\bf Figure 2} shows the cumulative populations
of the hot and cold regions for four equally-spaced values of the borderline reflection
energies. If one is of order the mean particle energy of unity the differences in densities
and temperatures are scarcely noticeable above the fluctuations, of order $\sqrt{100}$.
This Knudsen gas problem is a good example of entropy loss using standard Newtonian
mechanics. There is a clearcut successful energy separation between the hot and
cold regions for this particular formulation of Maxwell's Demon.

There is no doubt that the separation of hot particles from cold demonstrated
here can be predicted by kinetic theory (balancing the different densities and
temperatures so that the left-to-right and right-to-left mass fluxes sum to zero).
The successful separation of the energy into hotter and colder spatial regions has
required nothing more than a pair of conditionals in the motion equations, which satisfy
the conservation of energy as well as mass. Though no work is done momentum is not
conserved due to the stationary positions of the walls at $x = 0 \ {\rm and} \
\pm 10$ and $y = \pm 5$. A dense fluid provides a real challenge to the Demon in
that heat conductivity works against the thermal separation he promotes.

\section{Failure of Maxwell's Demon for a Dense Fluid}

\begin{figure}
\includegraphics[width=2.5 in,angle=-90.]{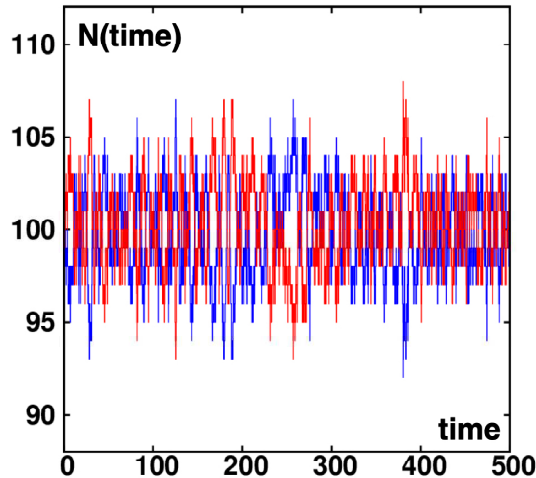}
\caption{
Time-dependence of the hot (red and in the right chamber) and cold (blue and at the left)
populations with a mirror kinetic energy of $0.2(E/N)$. Repulsive interactions between
pairs of particles are included here, increasing the mechanisms for heat transfer. As a
result Maxwell's Demon is relatively ineffective. The resulting systematic temperature
difference between the ``cold'' and ``hot'' regions is rather small. Just as in Figure 2
any particle crossing the center line $x=0$ is transmitted or reflected according to its
kinetic temperature.
}
\end{figure}

{\bf Figure 3} is taken from a simulation exactly similar to the Knudsen-gas model of
{\bf Figure 2} with the difference that the pair potential has been turned ``on''. Now
the particle energies are partly pair potential and the dynamics is no longer that of
an ideal gas. Exploration of such problems soon convinced us that the Demon was unable
to defeat the stabilizing energy flow due to heat conductivity enabled by pair-potential
interactions. We had not expected this problem though it is forecast by Gujarti in his
thermodynamic exploration of Maxwell's Demon\cite{b1}. Once the Demon's failure became
apparent in the dense-gas case we were at first unable to imagine and formulate a useful
Demon operating on individual particles. Instead we came to visualize two nonlocal Demons.
One of them operates solely left of center to lower the kinetic temperature there. The
``hot'' Demon operates at the right, raising that temperature, establishing a ``hot''
chamber. The two-Demon idea can be implemented with Nos\'e-Hoover dynamics, a stable
modification of Hamiltonian dynamics that makes it possible to specify a mean kinetic
energy imposed by integral feedback\cite{b3,b4,b5}. Specifying two temperatures is a
useful possibility. We explore that next.

\section{Hot and Cold Temperatures with two Thermostats}

To address Maxwell's problem with Nos\'e-Hoover thermal control we introduce {\it two}
new dynamical variables, $\zeta_{cold}$ and $\zeta_{hot}$. The equations of motion are then
modified for every particle at all times. This modification includes the influence of one
or the other of the two thermostat variables:
$$
x<0 \longrightarrow \dot p = F -\zeta_{cold}p \ ; \
x>0 \longrightarrow \dot p = F -\zeta_{hot}p \ . 
$$
The two friction coefficents obey feedback equations based on the two fluctuating kinetic
energies $K$ and their fixed target values $T$:
$$
\dot \zeta_{cold} = [(K/N)_{cold} - T_{cold}]/\tau^2 \ ; \
\dot \zeta_{hot}  = [(K/N)_{hot}  - T_{hot} ]/\tau^2 \ . 
$$
The relaxation times $\tau$ are free to choose. We set them equal to unity in the present
work. The two integral control variables $\zeta$ are the ``Nos\'e-Hoover Demons''.

{\bf Figure 4} shows the straightforward approach of 200-particle dynamics with
control of the two temperatures $T_c = 0.5$ and $T_h = 1.5$. Here no reflecting
mirror is used and all particles eventually sample the entire volume. Although
the dynamics is no longer Newtonian it does accomplish the goal of reducing the
entropy by extracting heat at the lower temperature and adding heat at the higher.
The entropy production,
$$
\dot S \equiv  (\dot Q/T_{cold}) - (\dot Q/T_{hot}), \ {\rm with} \ \dot Q>0,
$$
is extracted from the system and obviously inexorably reduces the entropy, eventually
to $-\infty$, and signals the formation of a fractal attractor within the 802-dimensional
phase space, which includes the two friction coefficients.  Though this modification of
Newtonian mechanics lacks the purity of Maxwell's Demon, who works with purely Newtonian
motion equations, it was our first attempt to extend the Demon idea to dense fluids.
Within 24 hours, on New Year's Eve, $2021 \longrightarrow 2022$, a better approach
suggested itself. We describe it next.

\begin{figure}
\includegraphics[width=1.8 in,angle=-90.]{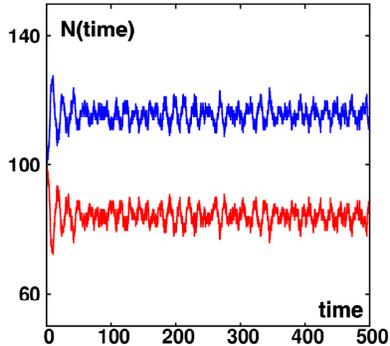}
\caption{
Time-dependence of the hot and cold populations with target temperatures of 0.5 and 1.5
using Nos\'e-Hoover thermostat variables. Repulsive interactions between pairs of
particles are included here, greatly increasing the heat conductivity. The
Nos\'e-Hoover Demons affect all particles simultaneously and maintain, along with
fluctuations, a systematic temperature difference between the ``cold'' and ``hot''
regions.  Just as in Figure 2 any particle crossing the center line $x=0$ is transmitted
or reflected according to its kinetic temperature, relative to unity.
}
\end{figure}

\pagebreak

\section{Maxwell's Demon for Soft-Disk Nonequilibrium Fluids}

How to approach the dynamic purity of Maxwell's Knudsen-gas Demon in a dense fluid ? First of
all, one needs an initial state from which a stationary, or time-periodic, nonequilibrium
system can result. Next, one needs a means, such as a Demon, for sustaining a distribution
different to any of Gibbs' equilibria, a state of less than maximum entropy.  Finally, one
needs a means of understanding the results of such a computation in thermodynamic or
hydrodynamic terms.

Pondering these needs led us to develop a dense-fluid molecular dynamics in which energy is
closely conserved through a Newtonian dynamics, with or without the actions of a Demon. A
simple task is reflecting particles at the system midline $x = 0$. Computations soon
revealed that equilibrium excursions from the mean number, 100, in the numbers of ``left''
and ``right'' particles were of order five. This work all used the short-ranged soft-disk
pair potential. By turning a diligent Demon ``on'' when the left and right chambers in a
purely-equilibrium simulation had populations 105 and 95, a nonequilibrium dense-fluid
soon develops. The two chambers do interact ``across'' the mid-line barrier which prevents
their mixing. See the left side of {\bf Figure 5} which shows the close correlation of the
left and right chamber temperatures. The interacting chambers evidently generate very
similar (no doubt identical) average temperatures despite their markedly different
densities (ten percent) and pressures, sustained by the rigid midline wall. The Demon's
dynamics for the situation shown in the figure is particularly simple. By remembering
the values of $x_i$ and $p_{x_i}$ from the previous timestep the Demon can simply set
the post-collision coordinate $x(t+dt) = x(t)$ and momentum, $p_x(t+dt) = -p_x(t)$. The
result conserves energy relatively well with $dt = 0.002$, ending up with an increase of
1/50 percent after five million timesteps. Such a simulation requires only about thirty
minutes on a laptop computer.

Evidently it is feasible to construct and maintain a nonequilibrium two-chamber system
so as to maintain thermal contact far from equilibrium by the diligent actions of a
Demon stationed at the wall common to both chambers. The Demon can easily be given
more complicated instructions, periodic in the time for example, without the need for
thermostats modifying Newton's laws of motion. We believe that this development sets
the stage for the simulation and analysis of a dynamics every bit as paradoxical as
Maxwell's Knudsen-gas idea. How this relates to the storage and erasure of information
is still a mystery to us. Perhaps a kind-hearted soul will consider such a problem and
provide an explanation which again saves the Second Law of Thermodynamics from Demons?

{\bf Figure 5} shows the development of the instantaneous energies in a system with
two equal-volume chambers containing 105 and 95 soft disks. At the left the mean
(cumulative) kinetic temperatures in the two chambers are shown. It is noteworthy that
the mean kinetic energy, $K/200$ has relatively small fluctuations compared to those
of the two chambers.

\begin{figure}
\includegraphics[width=1.6 in,angle=-90.]{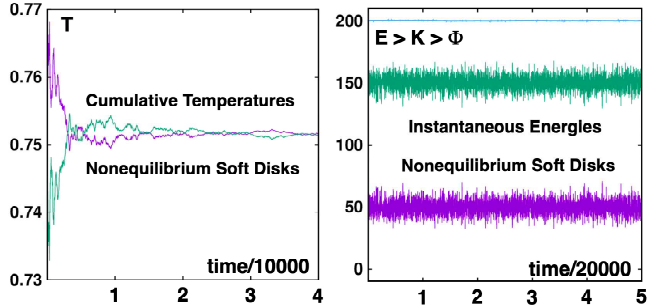}
\caption{
Energy relationships for a two-chamber nonequilibrium soft-disk fluid with densities
of 1.05 and 0.95. At the left we see that the time-averaged temperatures of the two
chambers match. At the right we see that despite the density and pressure differences
between the two chambers the nonequilibrium energy, $E = N = 200$, plotted at the top
is a constant of the motion. The dense-fluid Demon in this case causes all particle
collisions with the mid-line barrier at $x=0$ to be instantaneous and energy conserving.
}
\end{figure}

\pagebreak

\section{Summary}

The implementation of Maxwell's Demon using constrained versions of molecular dynamics
is straightforward. Any small energy errors due to a finite timestep (0.01 and 0.002 for the
fourth-order Runge-Kutta integrator used here) can be offset by a rescaling of the
particle velocities at the end of each timestep. With an exact interpolative treatment of
the mid-line collisions, the fourth-order Runge-Kutta errors can be reduced 16-fold by
simply halving the timestep.

Much has been written about the increase in entropy which must somehow compensate for
the lost entropy observed in the dynamics of thermal separation, where the entropy is
of order $N_{cold}\ln T_{cold} + N_{hot}\ln T_{hot}$ and is obviously maximized for the
case in which the numbers and temperatures in the two compartments are identical. Though
our dense-fluid model with two interacting compartments does come to thermal equilibration
the pressures and densities in the chambers necessarily differ from equilibrium states and
so indicate an entropy loss.

The computational work involved in the models introduced here is hardly different for the
equilibrium situation in which crossings of the midline are entirely ignored.  A
discussion of the Second Law for this model would likely note that a tremendous amount
of mental and numerical work needs to be carried out for any simulation, at or away from
equilibrium. Evidently the amount of this work has nothing to do with the mechanical
work envisioned in the First Law of Thermodynamics or the reversible heat transfers
visualized in the Second Law.

\pagebreak

\section{Acknowledgments}

We thank Professor Gujrati for sharing his work with us prior to its publication.
Brad Holian kindly reminded us of his own seminal work on two-chamber problems using
thermostats to impose nonequilibrium gradients\cite{b6,b7}.

\begin{figure}
\includegraphics[width=2 in,angle=0.]{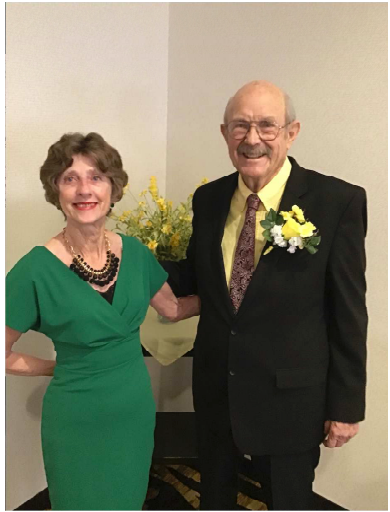}
\caption{
Bill and Carol Hoover at the wedding of Carol's Brother Larry Griswold to Kim Moore in
Colorado Springs, just before Christmas 2021. Bill and Carol moved to Ruby Valley Nevada
from California in 2005, after decades of work together at the Lawrence Livermore National
Laboratory.  They have published four books since their move to the Silver State:
{\it Time Reversibility, Computer Simulation, Algorithms, Chaos}; {\it Smooth Particle
Applied Mechanics---The State of the Art}; {\it Simulation and Control of Chaotic
Nonequilibrium Systems}, and {\it Microscopic and Macroscopic Simulation Techniques:
Kharagpur Lectures}. They are currently working with Clint Sprott on {\it Elegant Simulations},
scheduled for completion in 2022.
}
\end{figure}

\pagebreak


\begin{thebibliography}{99}

\bibitem{b1}  Puru Gujrati, ``Maxwell's Demon Must Remain Subservient to
              Clausius' Statement'', arXiv 2112.12300 (December, 2021).

\bibitem{b2}  W. G. Hoover and H. A. Posch, ``Large-System Hydrodynamic Limit'',
              Molecular Physics Report {\bf 10}, 70-85 (1995).

\bibitem{b3}  W. G. Hoover, ``Canonical Dynamics; Equilibrium Phase-Space
              Distributions'', Physical Review A {\bf 31} 1695-1697 (1985).

\bibitem{b4}  S. Nos\'e, ``A Molecular Dynamics Method for Simulations in the
              Canonical Ensemble'', Molecular Physics {\bf 52}, 255-268 (1984).

\bibitem{b5}  S. Nos\'e,``A Unified Formulation of the Constant Temperature
              Molecular Dynamics Methods'', Journal of Chemical Physics {\bf 81},
              511-519 (1984).

\bibitem{b6}  B. L. Holian, ``Simulations of Vibrational Relaxation in Dense
              Molecular Fluids. II. Generalized Treatment of Thermal Equilibration
              Between a Sample and a Reservoir'', Journal of Chemical Physics
              {\bf 117}, 1173-1180 (2002).

\bibitem{b7}  B. L. Holian ``Evaluating Shear Viscosity: Power Dissipated {\it Versus}
              Entropy Produced'',  Journal of Chemical Physics {\bf 117}, 9567-9568 (2002).

\end{thebibliography}
\end{document}